\documentclass[12pt]{article}

\usepackage{epsfig}
\usepackage{amssymb}
\usepackage{amsmath}
\usepackage{textcomp}

\newcommand{\Li}{\mbox{Li}_2}

\newcommand{\1}{1 \hspace{-1.05mm} {\rm l}}
\newcommand{\slp}{p \hspace{-1.9mm} /}
\newcommand{\slk}{k \hspace{-1.9mm} /}
\newcommand{\pf}{\bf \protect \boldmath}
\newcommand{\permille}{\mbox{\textperthousand}}

\newenvironment{Eqnarray}{\arraycolsep 0.14em \begin{eqnarray}}{\end{eqnarray}}
\newenvironment{Eqnarray*}{\arraycolsep 0.14em \begin{eqnarray*}}
{\end{eqnarray*}}
\renewcommand{\theequation}{\mbox{\arabic{equation}}}
\newcounter{saveeqn}
\newcommand{\alpheqn}{\setcounter{saveeqn}{\value{equation}}
\stepcounter{saveeqn}\setcounter{equation}{0}%
\renewcommand{\theequation}{\mbox{\arabic{saveeqn}\alph{equation}}}}
\newcommand{\reseteqn}{\setcounter{equation}{\value{saveeqn}}%
\renewcommand{\theequation}{\mbox{\arabic{equation}}}}

\begin{document}
\thispagestyle{empty}
\begin{flushright}
        MZ-TH/02-14\\
\end{flushright}
\vspace{0.5cm}
\begin{center}
 {\Large\pf $ O(\alpha_s) $}
 {\Large\bf  corrections to polarized top quark decay into a charged Higgs }
 {\Large\pf $ t(\uparrow) \rightarrow H^+ + b $}\\[7mm]
 {\large A.~Kadeer, J.G.~K\"orner and M.C.~Mauser}\\[13mm]
         Institut f\"ur Physik, Johannes Gutenberg-Universit\"at\\[2mm]
         Staudinger Weg 7, D-55099 Mainz, Germany\\[25mm]
\end{center}


\begin{abstract}
 \noindent We calculate the $ O(\alpha_s) $ radiative corrections to
 polarized top quark decay into a charged Higgs boson and a massive bottom quark in
 two variants of the two-Higgs-doublet model.
 The radiative corrections to the polarization asymmetry
 of the decay may become as large as $ 25 \, \% $.
 We provide analytical formulae for the unpolarized and
 polarized rates for $ m_b \neq 0 $ and for $ m_b = 0 $.
 For $ m_b = 0 $ our closed-form expressions for the
 unpolarized and polarized rates become rather compact.
\end{abstract}

\newpage


\section{Introduction}

The purpose of this paper is the evaluation of the first order QCD corrections
to the decay of a polarized top quark into a charged Higgs boson and a bottom
quark. 
Highly polarized top quarks will become available at hadron colliders
 through single top production, which occurs at the $ 33 \, \% $ level of the
 top quark pair production rate~\cite{MaPa97,EsMa02}.
 Future $ e^{+} e^{-} $ -- colliders will also be copious sources of polarized
 top quark pairs. The polarization of these can easily be tuned through the
 availability of polarized beams~\cite{PaSh96,KoMePr02}.
 Measurements of the decay rate and the polarization asymmetry in the decay
 $ t(\uparrow) \rightarrow H^{+} + b $ will be important for future tests
 of the Higgs coupling in the minimal supersymmetric standard model (MSSM).

 The $ O(\alpha_s) $ corrections to the decay rate $ t \rightarrow  H^{+} + b $
 have been calculated previously in 
\cite{Czar93(1),Czar93(2),Lial92,LiYa92,RTLS91,LiYu90,LiYu90-2,LiYa90,LiYa90-2}, 
and have been found to be important.
 The present paper provides the first calculation of the $ O(\alpha_s) $
 radiative corrections to the polarization asymmetry in polarized top quark decay
 $ t(\uparrow) \rightarrow H^{+} + b $.
 Depending on the value of $ \tan \beta $ and the mass of the charged Higgs boson
the radiative corrections to the
 polarization asymmetry  can become quite large (up to $ 25 \, \% $) in one
 variant of the two-Higgs-doublet model and must
 therefore be included in a decay analysis.

 The decay $ t \rightarrow H^{+} + b $ is analyzed in
 the rest frame of the top quark (see Fig.~1).
 The three--momentum $ \vec{q} $ of the $ H^+ $ boson
 points into the direction of the positive $ z $--axis.
 The polar angle $ \theta_P $ is defined as the angle
 between the polarization vector $ \vec{P} $ of the
 top quark and the $ z $--axis.

\begin{figure}[!h]        
 \begin{center} 
  \includegraphics[width=50mm]{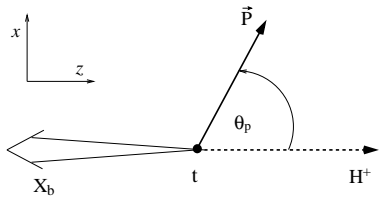}
  \caption{\sf \small Definition of the polar angle $ \theta_P $.
  $ \vec{P} $ is the polarization vector of the top quark.}
 \end{center}
\end{figure}
 In terms of the unpolarized rate $ \Gamma $ and the polarized
 rate $ \Gamma^P $ the differential decay rate is given by

 \begin{equation} 
  \label{difrate} 
  \frac{d \Gamma}{d \! \cos \theta_P} =
  \frac{1}{2} \, \Big(\Gamma + \mbox{P} \, \Gamma^P \cos \theta_P \Big) =
  \frac{1}{2} \, \Gamma \Big(1 + \mbox{P} \alpha_H \cos \theta_P \Big),
 \end{equation}

 \noindent where $P$ is the degree of polarization. 
 The polarization asymmetry $ \alpha_H $ is
 defined by $ \alpha_H = \Gamma^P / \Gamma $.
 Alternatively one can define a forward-backward asymmetry
 $ A_{F \! B} $ by writing
             
 \begin{equation} 
  \label{asymm}   
   A_{F \! B} = \frac{\Gamma_F - \Gamma_B}{\Gamma_F + \Gamma_B},
 \end{equation}

 \noindent where $ \Gamma_F $ and $ \Gamma_B $ are the
 rates into the forward and backward hemispheres, respectively.
 The two measures are related by
 $ A_{F \! B} = \frac{1}{2} \mbox{P} \alpha_H $.
 In our numerical results we shall always
 set $ \mbox{P} = 1 $ for simplicity.

 Technical details of our calculation can be found in \cite{FGKM02}.
 As in \cite{FGKM02} we use dimensional regularization
 ($ D = 4 \!-\! 2 \, \omega $ with $ \omega \ll 1 $) to
 regularize the ultraviolet divergences of the virtual corrections.
 We regularize the infrared divergences in the virtual one--loop and the 
real gluon emissions (tree--graph) 
 corrections by introducing a finite (small) gluon mass $ m_g \neq 0 $
 in the gluon propagator.
 In the tree--graph corrections the phase space boundary becomes deformed
 away from the IR singular point through the introduction of a (small) gluon mass.
 The logarithmic gluon mass dependence resulting from the regularization
 procedure cancels out when adding the virtual and tree--graph
 contributions.
 We have checked consistency with the Goldstone boson equivalence
 theorem which, in the limit $ m_{W^+} / m_t \rightarrow 0 $ and
 $ m_{H^+} / m_t \rightarrow 0 $, relates $ \Gamma $ and $ \Gamma^P $
 for $ t \rightarrow H^{+} +b $ to the unpolarized and polarized
 longitudinal rates $ \Gamma_L $ and $ \Gamma_L^P $ in the decay
 $ t \rightarrow W^{+} + b $ calculated in \cite{FGKM02}.


\section{Born term results}

 The coupling of the charged Higgs boson to the top and bottom quark in the
 MSSM can either be expressed as a superposition of scalar and pseudoscalar
 coupling factors or as a superposition of right-- and left--chiral coupling
 factors. The Born term amplitude is thus given by

 \begin{equation} 
  {\cal M}_0 = \bar{u}_b (a \1 + b \gamma_5) u_t =
  \bar{u}_b \left\{
  g_t \frac{\1 + \gamma_5}{2} +
  g_b \frac{\1 - \gamma_5}{2} \right\} u_t,
 \end{equation}

 \noindent where $ a \!=\! \frac{1}{2} (g_t + g_b) $ and
 $ b \!=\! \frac{1}{2} (g_t - g_b) $.
 The inverse relation reads $ g_t = a + b $ and $ g_b = a - b $.

 If the top quark mass $m_{t}$, bottom quark mass $m_{b}$ and the 
charged Higgs boson mass $m_{H^{+}}$ satisfy $ m_t > m_{H^+} + m_b$,
the top quark decay $ t\to H^{+} + b $ is possible in the MSSM.
 In order to avoid flavor changing neutral currents (FCNC) the
 generic Higgs coupling to all quarks has to be restricted.
 In the notation of \cite{GuHaKaDa90} in model~1 the doublet
 $ H_1 $ couples to all bosons and the doublet $ H_2 $
 couples to all quarks. This leads to the coupling factors

 \alpheqn
 \begin{Eqnarray} 
  \fbox{model 1:} & &
  a = \frac{g_w}{2 \sqrt{2} m_W}
  V_{tb}(m_t - m_b) \cot \beta, \\[3mm] & &
  b = \frac{g_w}{2 \sqrt{2} m_W}
  V_{tb}(m_t + m_b) \cot \beta,
 \end{Eqnarray}
 \reseteqn

 \noindent where $ V_{tb} $ is the CKM-matrix element and
 $ \tan \beta = v_2 / v_1 $ is the ratio of the vacuum expectation values
 of the two electrically neutral components of the two Higgs doublets.
 The weak coupling factor $ g_w $ is related to the usual Fermi coupling
 constant $ G_F $ by $ g_w^2 = 4 \sqrt{2} m _W^2 G_F $.

 In model~2, the doublet $ H_1 $ couples to the right--chiral
 down--type quarks and the doublet $ H_2 $ couples to the
 right--chiral up--type quarks. Model~2 leads to the coupling factors

 \alpheqn
 \begin{Eqnarray} 
  \fbox{model 2:} & &
  a = \frac{g_w}{2 \sqrt{2} m_W}
  V_{tb}(m_t \cot \beta + m_b \tan \beta), \\[3mm] & &
  b = \frac{g_w}{2 \sqrt{2} m_W}
  V_{tb}(m_t \cot \beta - m_b \tan \beta).
 \end{Eqnarray}
 \reseteqn

 Since $ m_b \ll m_t $ the $ b $ mass can always be safely neglected in model~1.
 In model~2 the left--chiral coupling term proportional to $ m_b \tan \beta $
 can become comparable to the right--chiral coupling term $ m_t \cot \beta $
 when $ \tan \beta $ becomes large.
 One cannot therefore naively set $ m_b = 0 $ in all expressions in model 2.
 One has to distinguish between the cases in which the scale of 
 $ m_{b} $ is set by $ m_t \cot^2 \beta $ and those in which the scale of
 $m_{b}$ is set by $ m_t $.
 In the former case one has to keep $m_{b}$ finite.
 In the latter case $m_{b}$ can safely be set to zero
 except for logarithmic terms proportional to $ \ln(m_b / m_t) $
 that appear in the NLO calculation.
 Keeping this distinction results in very compact closed--form
 expressions for the radiatively corrected unpolarized and polarized
 rates which we shall list in the main text.
 We shall, however, also present general $ m_b \ne 0 $
 results along with the $ m_b = 0 $ results.
 It is quite evident that one has to use the  $ m_b \ne 0 $ expressions
 for charged Higgs masses close to the top quark mass.
 
 For the amplitude squared one obtains
\begin{equation}
 \label{M2born}
|{\cal M}_{0}|^{2}= \sum_{s_{b}} {\cal M}_{0}^{\dagger} {\cal M}_{0}
=
 2(p_{t}\cdot p_{b})(a^{2}+b^{2})
 +2(a^{2}-b^{2})m_{t} m_{b}+4ab m_{t} (p_{b}\cdot s_{t})\,,
 \end{equation}
\noindent where
 $p_{t}$ and $p_{b}$ are the four--momenta of the top and bottom quarks. 
 $s_{t}$ denotes the polarization four--vector of the top quark.
 Accordingly the unpolarized and polarized Born term rates read

 \alpheqn
 \begin{eqnarray} 
  \label{born}    
  \!\Gamma_{Born}&\!=\!&
  \frac{m_t\sqrt{\lambda}}{16 \pi} \Big[
  (a^2 \!+\! b^2)(1 \!-\! y^{2}\!+\!\epsilon^{2}) + 2 (a^2 \!-\! b^2)\epsilon \Big], \\[2mm]
   \label{bornP} 
   \!\Gamma_{Born}^P &\!=\!& 
  \frac{m_t\sqrt{\lambda}}{16 \pi}\,
  2 a b \sqrt{\lambda}\,,
 \end{eqnarray}
 \reseteqn

 \noindent  where we have used the abbreviations
 $ \lambda = \lambda(1,y^2, \epsilon^2) $ with the K\"all\'en function  
  $ \lambda(a,b,c) := a^2 + b^2 + c^2 - 2 (a \, b + b \, c + c \,a)$.
 The scaled masses $y$ are $\epsilon$  are defined as
 \begin{equation} 
  y := \frac{m_{H^+}}{m_t}, \qquad
  \epsilon := \frac{m_b}{m_t}\,.
 \end{equation}
 As mass values we take $m_{b}=4.8\, {\rm GeV}$ and $m_{t}=175\, {\rm GeV}$ such
 that $\epsilon =0.027$. For later use we register that the factor 
 $ \mbox{PS}_2 = \sqrt{\lambda}/(16 \pi m_t)  $ is a two-body phase space factor.

 In the limit $ m_b \rightarrow 0 $ the unpolarized
 and polarized Born term rates simplify to

 \alpheqn
 \begin{eqnarray} 
  \label{bornzerobmass1} 
  \lim\limits_{m_{b}\to 0} \Gamma_{Born} & \!=\! &
  (a^2 + b^2) \big\{ 1 + \frac{a^2 - b^2}{a^2 + b^2}
  \frac{2 \epsilon}{1 - y^2} \big\} \, \hat{\Gamma}, \\[2mm]
\label{bornzerobmass2} 
\lim\limits_{m_{b}\to 0}\Gamma_{Born}^P&\!=\! & 2  a  b \, \hat{\Gamma},
 \end{eqnarray}
 \reseteqn
 where
 \begin{equation}
 \label{gammahat}
 \hat{\Gamma} = \frac{m_t(1 - y^2)^2}{16 \pi}\,.
 \end{equation}
 
 Eqs.~(\ref{bornzerobmass1}) and (\ref{bornzerobmass2}) are quite useful for a 
 discussion of the qualitative behavior of the unpolarized and polarized rates 
 in the two models as long as one is not too close to the kinematical limit
 $y=1- \epsilon$. In fact, for most of the $y$--region away from the 
 endpoint $y=1- \epsilon$, Eqs.~(\ref{bornzerobmass1}) and (\ref{bornzerobmass2}) 
 are an excellent approximation to Eqs.~(\ref{born}) and (\ref{bornP}).

\begin{figure*}[t]
\begin{center}
 \epsfig{figure=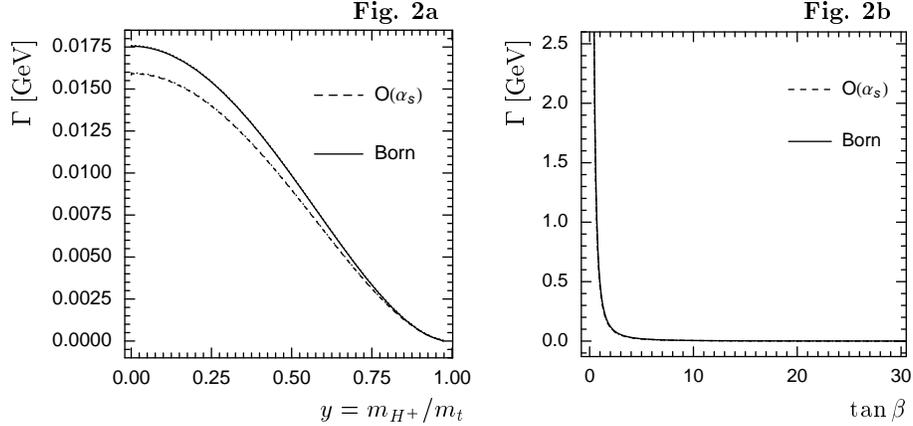,width=120mm}
\caption{\sf \small Unpolarized decay rate for model 1 as a 
  function of $y=m_{H^+}/m_t$ 
  (Fig.~ 2a; $ \tan \beta = 10 $) and as a function of $ \tan \beta $
  (Fig.~2b; $ m_{H^+} = 120\, {\rm GeV} $).
  Other parameters are
  $ m_b = 4.8\, {\rm GeV} $, $ m_t = 175 \,{\rm GeV} $.
  The barely visible dotted lines show the corresponding
  $ m_b \rightarrow 0 $ curves. }
\end{center}     
\end{figure*}

 For model 1 one has $(a^2-b^2)/(a^2+b^2)=-2\epsilon/(1+\epsilon^{2})$ and thus 
 the second term in 
 the curly brackets of Eq.~(\ref{bornzerobmass1}) can be safely dropped for 
 $\epsilon \to 0$ except when $y$ gets close to $1 - \epsilon$. For model 2 one has
 $(a^2-b^2)/(a^2+b^2)= 2\epsilon /(\cot^2 \beta+ \epsilon^2 \tan^2\beta)$, which,
 when combined with the factor $2 \epsilon$ in Eq.~(\ref{bornzerobmass1}), 
 can become as large as ${\cal O}(5 \%)$ for e.g. $\tan \beta \approx 5$. Since 
 this is further multiplied by $(1-y^{2})^{-1}$ we 
 shall therefore always keep the second term in the curly brackets of 
 Eq.~(\ref{bornzerobmass1}) in model 2 when making use of the $ m_b \rightarrow 0 $
 approximation. 

 In model 1 the $y$ dependence of the unpolarized and polarized rate is 
 essentially determined by the $(1-y^2)^2$ dependence of the overall
 factor $ \hat{\Gamma} $ in Eqs.~(\ref{bornzerobmass1}) and (\ref{bornzerobmass2}). 
 The overall scales of both the unpolarized and polarized rates are set by 
 $\cot^2 \beta$, which means that their ratio, the polarization asymmetry $\alpha_{H}$,
 is independent of the value of $\tan\beta$.

\begin{figure}[b] 
\begin{center}
  \epsfig{figure=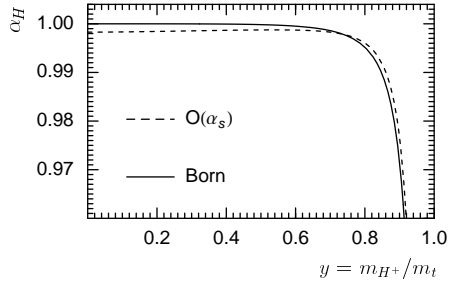,width=60mm}
\caption{\sf \small Polarization asymmetry $ \alpha_H $ for model~1 with
  $ m_b = 4.8 \,{\rm GeV} $ and $ m_t = 175 \,{\rm GeV} $
  as a function of $ m_{H^+}/m_t $. No value of $ \tan \beta $ is given since
  $ \alpha_H $ does not depend on $ \tan \beta $ in model 1. }
\end{center}
 \end{figure} 
\begin{figure*}[t] 
\begin{center}
 \epsfig{figure=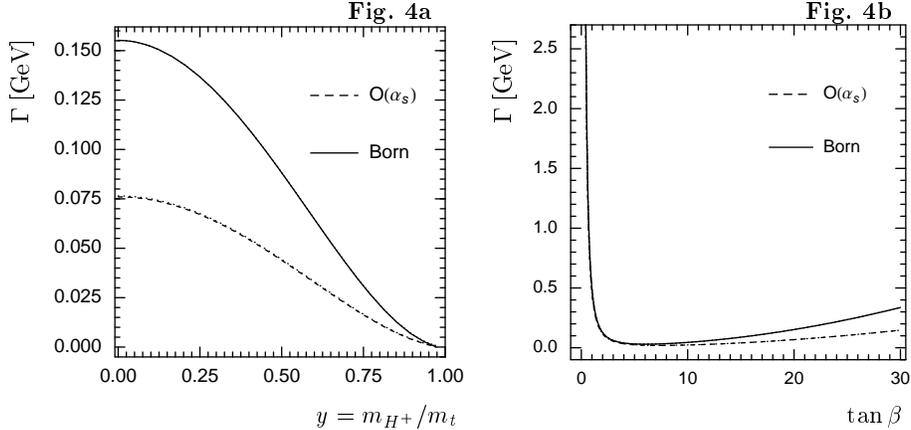,width=120mm}
\caption{\sf \small Unpolarized decay rate for model 2 as a function of $y=m_{H^+}/m_t$ 
  (Fig.~ 4a; $ \tan \beta = 10 $) and as a function of $ \tan \beta $
  (Fig.~4b; $ m_{H^+} = 120 \,{\rm GeV} $). Other parameters are
  $ m_b = 4.8 \,{\rm GeV} $, $ m_t = 175 \,{\rm GeV} $.
  The barely visible dotted lines show the corresponding
  $ m_b \rightarrow 0 $ curves.   }
\end{center}     
\end{figure*}
In Fig.~2a we show the unpolarized
 rate as a function of $y=m_{H^{+}}/m_{t}$ for $\tan \beta = 10$, which 
 exhibits the $(1-y^2)^2$ dependence of $ \hat{\Gamma} $ in Eq.~(\ref{gammahat}). 
 Compared to the partial Born term rate $\Gamma_{t\to W^{+}+b}=1.56 \,{\rm GeV} $
 the rate into a charged Higgs is generally quite small except for small $\tan\beta$ values.
This can also be seen in Fig.~2b where we plot the rate as a function of $\tan \beta$ for a 
 sample value of $m_{H^{+}}= 120\, {\rm GeV}$, which
 shows the $1/\tan^{2} \beta$ fall-off behavior of the overall factor 
 $(a^{2}+b^{2})\propto 1/\tan^{2} \beta $ in the rate expression 
 Eq.~(\ref{bornzerobmass1}). The rate into a charged Higgs can be seen to be quite 
 small for $\tan \beta$ values exceeding $\tan \beta=2$. One finds equality of 
 the partial rates into a $W^{+}$ and $H^{+}$ only at $\tan \beta=0.56$
 for $m_{H^{+}}= 120\, {\rm GeV}$. 
Such a small $\tan\beta$ value is excluded by the indirect limits in the  
$(m_{H^{\pm}},\tan\beta)$ plane~\cite{pdg06}.

In Fig.~3 we display the behavior of the polarization asymmetry $\alpha_H$
 as a function of $y=m_{H^{+}}/m_{t}$. As expected from the flat behavior of
 $(a^{2}+b^{2})/(2ab)$ the polarization asymmetry $\alpha_H$ stays very close to its maximal value 
 $\alpha_H \approx 1$ except for the endpoint region $y \to (1- \epsilon)$. At 
 $y=1- \epsilon$ the polarization asymmetry drops to $\alpha_H = 0$ (not shown).
 We do not show a plot of the 
 polarization asymmetry as a function of $\tan \beta$, because, as mentioned
 before, the polarization asymmetry does not depend on $\tan \beta$ for model 1.

\begin{figure*}[t] 
\begin{center}
 \epsfig{figure=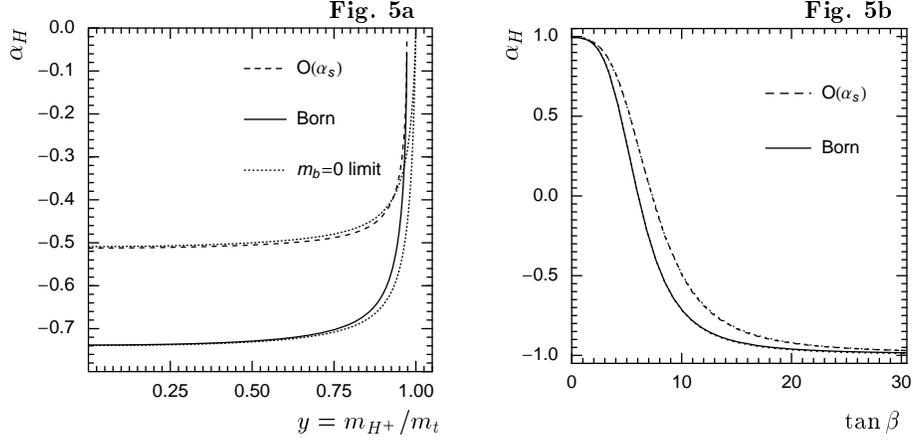,width=120mm}
\caption{\sf \small Polarization asymmetry  $ \alpha_H $ for model~2 with
  $ m_b = 4.8 \,{\rm GeV} $ and $ m_t = 175 \,{\rm GeV} $
  as a function of
  $ m_{H^+}/m_t $ (Fig.~5a, $ \tan \beta = 10 $) and as function of
  $ \tan \beta $ (Fig.~5b, $ m_{H^+} = 120 \,{\rm GeV} $).
  The dotted lines (barely visible in Fig.~5b) show the corresponding
  $ m_b \rightarrow 0 $ curves.  }
\end{center}     
\end{figure*}

The $y$ and $\tan \beta$ dependence of model 2 is somewhat more involved.
 The overall scales for the unpolarized and polarized rates are now set by
 $(\cot^2 \beta \pm \epsilon^2 \tan^2 \beta)$, respectively. The $y$ dependence 
 of both the unpolarized and polarized rates is again
 close to the $\epsilon \to 0$ limit 
since they are dominated by the $(1-y^2)^2$ behavior of $ \hat{\Gamma} $.
 However, in the case of the unpolarized rate, the $y$ dependence is slightly modified by the
 curly brackets in Eq.~(\ref{bornzerobmass1}). In Fig.~4a we display the dependence
 of the rate on $y=m_{H^{+}}/m_{t}$ where we take again $\tan \beta = 10$ as 
 a numerical example. As in model 1 the functional behavior is again essentially 
 determined by the overall $(1-y^2)^2$ behavior of Eq.~(\ref{bornzerobmass1})
 except for the region close to the endpoint where the second term in the
 curly brackets of Eq.~(\ref{bornzerobmass1}) comes into play. The rate is largest for 
 $m_{H^{+}}= 0$ and drops to zero towards the phase space boundary $y=1-\epsilon$.
The model~2 rate is considerably larger than the model~1 rate, depending, of course
on the value of $\tan\beta$.
This is because of the overall rate
factor $(a^{2}+b^{2})$ whose model~2/model~1 ratio
\begin{equation}
\frac{(a^{2}+b^{2})|_{\rm model\,2}}{(a^{2}+b^{2})|_{\rm model\,1}}
=\frac{(1+\epsilon^{2}\tan^{4}\beta)}{(1+\epsilon^{2})}
\end{equation}
is larger than 1 for $\tan\beta >1$.
For example, for $\tan\beta=10$ and $\epsilon=0.027$
one has $(a^{2}+b^{2})|_{\rm model\,2}/(a^{2}+b^{2})|_{\rm model\,1}=8.28$.
The second term in the curly brackets of Eq.~(\ref{bornzerobmass1}) gives a much smaller 
enhancement. For example, for $y=0$, $\tan\beta=10$ and $\epsilon=0.027$,
the enhancement due to the second term in the curly brackets of Eq.~(\ref{bornzerobmass1})
is only 3.5\%.
Nevertheless, the model~2 rate is still small compared to the Standard Model Born term
rate $\Gamma_{t\rightarrow W^{+}+b}$=1.56~GeV for most of the range of $\tan\beta$ value.

 In Fig.~4b we plot the $\tan \beta$ dependence of the unpolarized rate for 
 $m_{H^{+}}= 120 \,{\rm GeV}$. The $ \tan \beta $ dependence
 of the rate is determined by the overall 
 $ \tan \beta $--dependent factor 
 $(a^{2}+b^{2})\propto (1 + \epsilon^{2}\tan^4 \beta)/\tan^2 \beta$ with a slight
 modification due to the second term in the curly brackets in 
 Eq.~(\ref{bornzerobmass1}). As in model~1 the rates drop very fast at the origin for 
 small $\tan \beta$ values due to the dominant $1/\tan^2 \beta$ behavior of the 
 overall factor $(a^{2}+b^{2})$. Contrary to model 1 the rate rises quadratically  for larger 
 $\tan \beta$ values after going through a minimum at $\tan \beta \approx 7 $
 due to the $\tan^2 \beta$ term in the overall rate factor. It reaches equality
 with the Standard Model Born term rate $\Gamma_{t\to W^{+}+b}=1.56\,{\rm GeV} $ 
 at $\tan \beta= 0.56$ and $\tan \beta=64.83$. Again these values of $\tan\beta$ 
are excluded by the indirect limits in the $(m_{H^{\pm}},\tan\beta)$ plane~\cite{pdg06}.

All of what has been stated in the last paragraph does not change qualitatively but
quantitatively if one uses a running mass for the bottom quark instead of a fixed pole mass.
At one--loop running one would then have $\bar{m}_{b}(m_{t})=1.79$~GeV
and $\epsilon=\bar{m}_{b}(m_{t})/m_{t}=0.010$ instead of $\epsilon=0.027$.
The main effect would be a reduction of the model~2 rate for larger values of $\tan\beta$
due to the overall factor $(a^{2}+b^{2})$. For example, for $\tan\beta=10$
one would have a rate reduction by a factor of 0.24 due to the factor $(a^{2}+b^{2})$
when one uses a running bottom quark mass instead of a fixed pole mass.

 The main qualitative features of the behavior of the polarization asymmetry
 $ \alpha_H $ in model~2 can again be understood from the $ m_b \rightarrow 0 $ 
 behavior of $ \alpha_H $ given by the ratio of Eq.~(\ref{bornzerobmass2}) and
 Eq.~(\ref{bornzerobmass1}). One has
 \begin{equation}
\label{asym}
\lim\limits_{m_{b}\to 0}\alpha_H = \frac{1-\epsilon^2 \tan^4\beta}{1+\epsilon^2 \tan^4\beta}
\left\{ 1+ \frac{4 \epsilon}{(1-y^{2})}\,\, \frac{\epsilon \tan^2\beta}
{1+\epsilon^2 \tan^4\beta} \right\}^{-1}\,.
\end{equation} 
 The $y$ dependence comes only from the second term in the curly brackets of
 Eq.~(\ref{asym}) and is therefore not very pronounced. This can be seen in
 Fig.~5a, where we show a plot of the polarization asymmetry $ \alpha_H $ as a 
 function of $y$ for $ \tan \beta =10 $ in model~2. $ \alpha_H $ 
 is large and negative with only little dependence on the Higgs mass
 except for the region close to the phase space boundary $y=1-\epsilon$ where 
 $ \alpha_H $ approaches zero. $ \alpha_H $ is negative due to the fact that 
 the numerator factor $(1 - \epsilon^{2}\tan^4 \beta)$ is negative for 
 $\tan \beta = 10$.  For the 
 smaller $y$ values the functional behavior is dominated by the first term 
 in the curly brackets of Eq.~(\ref{asym}), and therefore the curve is rather flat. 
 Towards $y=1-\epsilon$ the second term in the curly bracket
 of Eq.~(\ref{asym}) becomes dominant and the curve bends up
 and finally reaches zero at $y=1-\epsilon$. The $\epsilon=0$ approximation 
 can be seen to be quite good up to $y \approx 0.7$.
 In Fig.~5b we show a plot of the polarization asymmetry $ \alpha_H $ as a function of
 $ \tan \beta $ for $m_{H^{+}}= 120\, {\rm GeV}$, i.e. $y=0.69$. 
 Contrary to model~1, $\alpha_H $ shows 
 a strong dependence on the value of $ \tan \beta $. $\alpha_H $ is 
 positive/negative for small/large values of $ \tan \beta $ and goes through
 zero for $ \tan \beta = \sqrt{m_t / m_b}=6.04$ close to where the rate has a 
 minimum.  Beyond the zero position $ \alpha_H $ rapidly approaches
 values close to $ \alpha_H = - 1 $. Both Figs.~5a and 5b show that  
 $m_{b} \to 0$ is a very good approximation. At the scale of the figures the 
 dotted curve cannot be discerned from the exact full curves.


\section{Virtual corrections}

 The virtual one--loop corrections to the
 $ (tH^{+}b) $--vertex exhibit both
 IR-- and UV--singularities.
 As mentioned before, the UV--singularities are
 regularized in $ D = 4 \!-\! 2 \, \omega $ dimensions,
 whereas the IR--singularities are regularized by a small
 gluon mass $ m_g $. The scaled gluon mass is denoted by 
\begin{equation}
\Lambda:=\frac{m_{g}}{m_{t}}\,.
\end{equation}
 The renormalization of the UV--singularities
 is done in the ``on--shell'' scheme.

 The renormalized amplitude of the virtual corrections in
 the right-- and left--chiral representation can be written as

 \begin{equation} 
  {\cal M}_{loop} = \bar{u}_b \Big\{
  \Big( a \1 + b \gamma_5  \Big) \Lambda_1 +
  a \Lambda_2 + \delta \Lambda \Big\} u_t,
 \end{equation}

 \noindent where the functions $ \Lambda_1 $ and $ \Lambda_2 $ read
($C_{F}=\frac{4}{3}$)

 \alpheqn
 \begin{Eqnarray} 
  \Lambda_1 & = &\frac{\alpha_{s}}{2\pi} C_F \bigg\{ ( m_{t}^{2}+m_{b}^{2}-m_{H}^{2}) \, C_0
  - \Big[2 m_{t}^{2}-m_{H}^{2} + m_b (m_t + m_b)\Big] \, C_1 \nonumber \\
&&\qquad\quad - \Big[2 m_{b}^{2}-m_{H}^{2} + m_t (m_t + m_b)\Big] \, C_2+ 
  2B_0 -1 \bigg\},\nonumber  \\[2mm]
  \Lambda_2 & = &\frac{\alpha_{s}}{2\pi}  C_F \, \Big\{
  2m_b m_t (C_1 + C_2) \Big\}.
 \end{Eqnarray}
 \reseteqn

\noindent The standard one--loop integrals $B_{0}$, $ C_0 $, $ C_1$ 
 and $C_{2}$ are given in Appendix~\ref{loopInt-t2Hb}.
 After a linear transformation these loop functions can be seen
 to agree with Eqs.~(8) and (9) in Ref.~\cite{Lial92}.
 Following Refs.~\cite{Czar93(2),Lial92,LiYu90,BrLe80}
 the counter term of the vertex is given by

 \begin{eqnarray} 
 \delta \Lambda &=&
  ( a + b ) \frac{\1 + \gamma_5}{2} \bigg(
  \frac{1}{2} (Z_2^t - 1) + \frac{1}{2} (Z_2^b - 1) -
  \frac{\delta m_t}{m_t} \bigg)\nonumber   \\[2mm]
 &&\,\, +  ( a - b ) \frac{\1 - \gamma_5}{2} \bigg(
  \frac{1}{2} (Z_2^t - 1) + \frac{1}{2} (Z_2^b - 1) -
  \frac{\delta m_b}{m_b} \bigg).
 \end{eqnarray}

 In the ``on--shell'' scheme the wavefunction renormalization constant
 $ Z_2^q $ and the mass renormalization constant $ \delta m_q $ can be
 calculated from the renormalized QCD self-energy $ \Sigma_q (p) $ of
 the quarks $ q = t,b $.
 The evaluation of the two conditions $ \Sigma_q (p) |_{p \!\!\!/ = m_q} = 0 $
 and $ \partial \Sigma_q (p) / \partial \slp |_{p \!\!\!/ = m_q} = 0 $ lead to
 the following renormalization constants

 \alpheqn
 \begin{Eqnarray} 
  && \hspace*{-5mm}Z_2^q =
  1 \!-\! \frac{\alpha_s}{4 \pi} C_F \left[
  \frac{1}{\omega} \!-\! \gamma_E \!+\!
  \ln \frac{4 \pi \mu^2}{m_q^2} \!+\!
  2 \ln \frac{m_g^2}{m_q^2} \!+\! 4 \right], \\[2mm]
  && \hspace*{-6mm}\delta m_q=
  \frac{\alpha_s}{4 \pi} C_F \, m_q \, \left(
  \frac{3}{\omega} - 3 \gamma_E +
  3 \ln \frac{4 \pi \mu^2}{m_q^2} + 4 \right),
 \end{Eqnarray}
 \reseteqn

 \noindent where $ p $ is the four--momentum and $ m_q $
 is the mass of the relevant quark.

 Putting everything together the virtual one--loop contributions to the unpolarized 
 and polarized rates read ($ \lambda = \lambda(1,y^2, \epsilon^2) $)

 \alpheqn
 \begin{Eqnarray} 
  \label{GloopA}  
  \Gamma_{loop} & = &
  \Gamma_{Born} \bigg( Z_2^t - 1 + Z_2^b - 1 -
  \frac{\delta m_t}{m_t} - \frac{\delta m_b}{m_b} +
  2 \Lambda_1 \bigg)  \nonumber \\ 
&&+  \frac{m_t\sqrt{\lambda}}{16 \pi } \nonumber \bigg(
  2 \, a^2 \, \Lambda_2 \left((1+\epsilon)^{2}-y^{2}\right) - 2ab \, 
  (1-y^{2}+\epsilon^{2})
  \bigg( \frac{\delta m_t}{m_t} -
  \frac{\delta m_b}{m_b} \bigg) \bigg), \\[2mm]
  \label{GloopB}
  \Gamma_{loop}^{P} & = &
  \Gamma_{Born}^P \bigg( Z_2^t \!-\! 1 + Z_2^b \!-\! 1 -
  \frac{\delta m_t}{m_t} \!-\! \frac{\delta m_b}{m_b} +
  2 \Lambda_1+\Lambda_{2} \bigg) \nonumber  \\[2mm] 
&& +
  \frac{m_t\sqrt{\lambda}}{16 \pi } (a^2 + b^2) \, \sqrt{\lambda}
  \bigg( \frac{\delta m_b}{m_b} -
  \frac{\delta m_t}{m_t} \bigg).
 \end{Eqnarray}
 \reseteqn

 The renormalized virtual one--loop correction to the unpolarized rate
 (\ref{GloopA}) is in agreement with Ref.~\cite{Lial92}.
 The result for the virtual one--loop correction to the polarized rate is new.
 Note that the infrared divergent terms residing in the renormalization factor 
 $ Z_2^q $
 and in the integral term $ C_0 $ in $ \Lambda_1 $ are proportional to the
 Born term rates $ \Gamma_{Born} $ and $ \Gamma_{Born}^P $, respectively.


\section{Tree--graph contributions}

 The ${\cal O}(\alpha_{s})$ real gluon emission (tree--graph) amplitude reads

 \begin{equation} 
  {\cal M}_{tree} = g_s \frac{\lambda^a}{2} \bar{u}_b \bigg\{
  \frac{2 p_t^{\sigma} - \slk \gamma^{\sigma}}{2 k \!\cdot\! p_t} -
  \frac{2 p_b^{\sigma} + \gamma^{\sigma} \slk}{2 k \!\cdot\! p_b} \bigg\}
\bigg( a \1 + b \gamma_5 \bigg) u_t \,\varepsilon_{\sigma}^{\ast}(k,\lambda)\,,
 \end{equation}

 \noindent where the  first and second terms in the curly brackets refer to
 real gluon emission from the top quark and the bottom quark, respectively.
The four--momenta of the charged Higgs and the gluon are denoted 
by $q$ and $k$. The polarization vector of the gluon with momentum $k$
and spin $\lambda$ is denoted by $\varepsilon(k,\lambda)$.

In the square of the tree--graph amplitude the terms without gluon momentum in the numerator, 
i.e. the terms proportional to
$p_{t}$ and $p_{b}$ in the numerator lead to IR--divergences, which are
regulated by a small gluon mass.
Separating these terms will split the squared tree--graph amplitude 
into an IR--convergent part $|{\cal M}_{tree}^{conv}|^2$ and an IR--divergent
part proportional to the soft gluon factor $|{\cal M}|^2_{SGF}$ as follows:
 \begin{equation} 
 \label{primary IR-isolation}
  |{\cal M}_{tree}|^2 =
   |{\cal M}_{tree}^{conv}|^2 + |\widetilde{\cal M}_{0}|^2 |{\cal M}|^2_{SGF}
 \end{equation}
where the factor $|\widetilde {\cal M}_{0}|^{2}$
and the universal soft gluon  (or eikonal) factor $|{\cal M}|^2_{SGF}$
are given by
 \begin{Eqnarray} 
\label{MsgfDef}
  |{\cal M}|^2_{SGF} &=& -
  4\pi \alpha_s  C_F \bigg[
  \frac{m_t^2}{(k \!\cdot\! p_t)^2} +
  \frac{m_b^2}{(k \!\cdot\! p_b)^2}  - 2  \frac{p_b \!\cdot\! p_t}
  {(k \!\cdot\! p_b)(k \!\cdot\! p_t)} \bigg]\,,\\[2mm]
\label{M2borntilde}
|\widetilde{\cal M}_{0}|^{2}&=& 2(p_{t}\cdot p_{b})(a^{2}+b^{2})
+2(a^{2}-b^{2})m_{t} m_{b}+4ab m_{t} (p_{b}\cdot s_{t})\,.
 \end{Eqnarray}
\noindent Note that the factor $|\widetilde{\cal M}_{0}|^2$ has the same 
analytical form as 
the squared Born term amplitude $|{\cal M}_{0}|^{2}$ listed in Eq.~(\ref{M2born})
but must of course be evaluated for $ p_t = p_b + q + k $,
i.e. one has $|\widetilde{\cal M}_{0}|^2 (k=0)=|{\cal M}_{0}|^{2}$.

When integrating $|\widetilde{\cal M}_{0}|^2 |{\cal M}|^2_{SGF}$
one has to take care of the fact that both terms in the product 
depend on the gluon momentum.
Since $(|\widetilde{\cal M}_{0}|^2 -|{\cal M}_{0}|^2) |{\cal M}|^2_{SGF} $
is convergent and thus can be integrated without a gluon mass regulator,
we further isolate the IR--divergent part by writing
\begin{equation}
\label{Mr}
  |{\cal M}_{tree}|^2 =
  \Big\{ |{\cal M}_{tree}^{conv}|^2 +
  \Big( |\widetilde{\cal M}_{0}|^2 - |{\cal M}_{0}|^2 \Big)
  |{\cal M}|^2_{SGF} \Big\}+ |{\cal M}_{0}|^2 |{\cal M}|^2_{SGF} ,
 \end{equation}
 where the gluon momentum dependence in the IR divergent last term of (\ref{Mr})
 solely resides in the soft gluon factor $|{\cal M}|^2_{SGF}$.

 The same universal soft gluon factor $|{\cal M}|^2_{SGF}$ appears in the 
 calculation of the
 radiative corrections to $ t \rightarrow W^{+} + b $.
 We can therefore take the result of its phase space integration
 (with $ m_g \ne 0 $) from Eq.~(63) in Ref.~\cite{FGKM02}.

 The IR--convergent part of $ |{\cal M}_{tree}|^2 $ is thus given by

\begin{Eqnarray} 
\label{Mrcon-all}
 |{\cal M}_{tree}^{conv}|^2 &+& \Big( |\widetilde{\cal M}_{0}|^2 -
  |{\cal M}_{0}|^2 \Big) |{\cal M}|^2_{SGF}=  8\pi \alpha_s  C_F \frac{k \!\cdot\! {q}}
  {(k \!\cdot\! p_t)(k \!\cdot\! p_b)} \bigg\{
  (a^2 + b^2) (k \!\cdot {q}) +
  \nonumber \\ 
& &  -2a b m_{t} \bigg[
  \bigg( \frac{m_t^2 \!+\! m_{H^+}^2 \!-\! m_b^2}{2 k \!\cdot\! p_t} \!-\! 1 \!-\!
  \frac{m_{H^+}^2}{k \!\cdot\! {q}} \bigg) (k \!\cdot\! s_t) \!+\! (q\!\cdot\! s_{t})\nonumber \\
& &  \!+\!\bigg( \frac{m_t^2}{ k \!\cdot\! p_t} \!-\!
  \frac{m_b^2}{ k \!\cdot\! p_b} \!-\!2  \!-\!
  \frac{m_{H^+}^2}{ k \!\cdot\!{q}} \bigg)\Big( (q \!\cdot\! s_t)\!-\!(q \!\cdot\! s_t)|_{k=0}\Big)
  \bigg] \bigg\}.
 \end{Eqnarray}

 It is quite remarkable that the unpolarized and polarized pieces of the
 convergent tree--level contribution are proportional to $ (a^2 + b^2) $
 and $ 2 a b $, respectively. 

 The phase space integration is done with respect to the gluon energy $ k_0 $
 and the $ H^{+} $ boson energy $  E_{H^+}$, where the $ k_0 $
 integration is done first. It is convenient to introduce the scaled invariant
 mass of the bottom quark-gluon system $z=\frac{(p_{b}+k)^{2}}{m_{t}^{2}}$.
 Using the relation $ E_{H^+} = (m_t^2 + m_{H^+}^2 - m_t^2 z)/(2 m_t) $ we
 change to the integration variable $z$.

Note that the last line of the Eq.~(\ref{Mrcon-all})
comes from the term 
$\Big( |\widetilde{\cal M}_{0}|^2 - |{\cal M}_{0}|^2 \Big)   |{\cal M}|^2_{SGF} $
in Eq.~(\ref{Mr}). The relevant scalar products are evaluated in the top quark 
rest frame. They read 

\begin{Eqnarray} 
(q \!\cdot\! s_t)&=&- \frac{1}{2}\sqrt{\lambda(1,y^{2},z)}\,m_{t}\cos\theta_{P}\,,\\
(q \!\cdot\! s_t)|_{k=0}&=&- \frac{1}{2}\sqrt{\lambda(1,y^{2},\epsilon^{2})}\,m_{t}\cos\theta_{P}\,.
\end{Eqnarray}
 
 After $k_0$ integration the remaining phase space integrations can
 be reduced to the following classes of basic 
integrals ($\lambda^{\prime}=\lambda(1,y^{2},z)$):

 \alpheqn
 \begin{Eqnarray} 
  \label{master}  
 && R(n)  :=  \int dz
    \frac{1}{(z - \epsilon^2) \sqrt{\lambda^{\prime n}}}, \\[2mm]
  && R(m,n)  :=  \int dz
    \frac{z^m}{\sqrt{\lambda^{\prime n}}}, \\[2mm]
  && S(n)  := \! \int\! dz
    \frac{1}{(z \!-\! \epsilon^2) \sqrt{\lambda^{\prime n}}} \ln \bigg[
    \frac{1 \!-\! y^2 \!+\! z \!+\! \sqrt{\lambda^{\prime}}}
    {1 \!-\! y^2 \!+\! z \!-\! \sqrt{\lambda^{\prime}}} \bigg] , \\[2mm]
  && S(m,n) := \! \int\! dz
    \frac{z^m}{\sqrt{\lambda^{\prime n}}} \ln \bigg[
    \frac{1 - y^2 + z + \sqrt{\lambda^{\prime}}}
    {1 - y^2 + z - \sqrt{\lambda^{\prime}}} \bigg] \, , 
 \end{Eqnarray}
\reseteqn 

\noindent with the integration limits 
$\epsilon^{2} \leqslant z \leqslant (1-y)^{2}$.
The basic integrals needed in the present application are listed in Appendix~\ref{basicInt}.

 Finally, the unpolarized and polarized tree--graph contributions read
 \alpheqn
 \begin{Eqnarray} 
 \Gamma_{tree}  &=& - \frac{1}{4 \pi m_t}\bigg[
  \frac{\alpha_s}{4 \pi} C_F \, m_t^2 
(a^2 + b^2)
  \bigg\{
  \frac{3}{4} R(0,-1) +\frac{\epsilon^2(1-y^2)}{4}R(-2,-1)\nonumber \\
&&-\frac{1-y^2+3\epsilon^2}{4}R(-1,-1)  +\frac{1}{2}\epsilon^2 S(0,0)-\frac{1}{2}S(1,0)
  \bigg\}
  +  \mbox{PS}_2^{-1} \, \Gamma_{Born} \, S(\Lambda)
  \bigg]\,,\nonumber \\
 \end{Eqnarray}
\noindent and 
 \begin{Eqnarray} 
 \Gamma_{tree}^{P} &=& - \frac{1}{4 \pi m_t}\bigg[
  \frac{\alpha_s}{4 \pi} C_F \, m_t^2\,   
2 a b \,
    \bigg\{
     -2 \sqrt{\lambda}\,R(-1) 
   + 2\lambda\,R(0)  + \frac{1}{4}{( 1 -y^2)}^2 \epsilon^2 R(-2,0)\nonumber \\
&& -  \frac{1}{4}\Big( { ( 1 - y^2 ) }^2
     + 2\,( 3 - y^2 ) \,\epsilon^2 \Big)\,R(-1,0)
     - \frac{1}{4} ( 2 + 10\,y^2 - \epsilon^2 ) \,R(0,0) +
\frac{7}{4}R(1,0) \nonumber \\
&& +  ( 1 - y^2 + \epsilon^2 ) \,\sqrt{\lambda}\,S(0)
  - ( 1 - y^2 + \epsilon^2 )\lambda \,S(1)
  + \sqrt{\lambda} \, S(0,0) + \frac{1}{2}\Big( 4\,y^2\,\left( 1 - y^2 \right)\nonumber \\
&&
  + \left(7 + 5\,y^2 \right) \,\epsilon^2 - 2\,\epsilon^4 \Big) \,
     S(0,1)- \frac{1}{2}\left( 3 - 3\,y^2 + \epsilon^2 \right) \,S(1,1) - \frac{1}{2}S(2,1) \bigg\} \nonumber \\
 &&+
  \mbox{PS}_2^{-1} \, \Gamma_{Born}^P \, S(\Lambda)
  \bigg]\,.
 \end{Eqnarray}
 \reseteqn

\noindent As before, $ \mbox{PS}_2 = \sqrt{\lambda}/16 \pi m_t $
 is the two--body phase space factor.
$S(\Lambda)$ is the integration of the soft gluon 
 factor $|{\cal M}|^{2}_{SGF}$ defined in Eq.~(\ref{MsgfDef}), i.e.   
\begin{equation}
       S(\Lambda) := (-4\pi
m_{t})\frac{1}{2m_{t}}\frac{1}{(2\pi)^{5}}\int \frac{d^{3}\vec{p}_{b}}{2E_{b}}
     \frac{d^{3}\vec{q}}{2E_{H}}
     \frac{d^{3}\vec{k}}{2E_{g}} \,\delta^{(4)}(p_{t}-q-p_{b}-k) |{\cal
M}|^{2}_{SGF}\,.\nonumber \\
     \end{equation}

The integration $S(\Lambda)$ was done e.g. in \cite{FGKM02}. For
 completeness we list the result in Appendix~\ref{SGF-int}.


\section{Results}
\label{sec-results}
For ease of comparison with the results of \cite{Czar93(1),Czar93(2)} 
we define the following abbreviations:
\begin{alignat}{2} 
   \hat{p}_0 &= \frac{1}{2}(1+\epsilon^2-y^2)\,, \qquad &\hat{w}_0 &= \frac{1}{2}(1-\epsilon^2+y^2)\,,\nonumber\\
 \hat{p}_3 &=\frac{1}{2}\sqrt{\lambda(1,\epsilon^2,y^2)}\,,\qquad  & \hat{w}_3 &=\hat{p}_{3}\,,\nonumber\\
  \hat{p}_{\pm} &= \hat{p}_0\pm \hat{p}_3 \,, \qquad  &\hat{w}_{\pm} &= \hat{w}_0\pm \hat{w}_3\,,\nonumber \\ 
Y_p &=\frac{1}{2}\ln\frac{\hat{p}_+}{\hat{p}_-}\,,\qquad  &Y_w &=\frac{1}{2}\ln\frac{\hat{w}_+}{\hat{w}_-}\,,
 \end{alignat}
where the hat symbols on the scaled momenta $\hat{p}_i$ and $\hat{w}_i$ accentuate
the fact that we are dealing with dimensionless quantities.
 The complete $ O(\alpha_s) $ results are obtained by summing the Born term,
 the virtual one--loop and the tree--graph contributions, i.e.
\begin{Eqnarray}
 \frac{d\Gamma}{d\cos\theta_{P}}&=&\frac{1}{2}(\Gamma + P \, \Gamma^{P}\cos\theta_{P} )\nonumber\\
&=& \frac{1}{2}
 \bigg[ \Big(\Gamma_{Born}+\Gamma_{NLO}\Big)+ \Big(\Gamma^{P}_{Born}+\Gamma^{P}_{NLO}\Big)P\cos\theta_{P}  \bigg].
\end{Eqnarray}

\noindent The unpolarized $\mathcal O(\alpha_{s})$ corrections are
given by\footnote{This result is written in the same form as in \cite{Czar93(2)}. } 
\begin{equation}
\label{t2Hbtotalunpol}
\Gamma_{NLO}=\Gamma_{loop}+\Gamma_{tree}=\frac{\alpha_{s}  }{8 \pi^{2}}C_{F}m_{t}\Big[
 (a^{2}\!+\!b^{2})\,  G_{+}+  (a^{2}\!-\!b^{2})\,\epsilon\,  G_{-} + a b\, G_{0}\Big]
\end{equation}
with the coefficient functions
\begin{Eqnarray}
G_{+} &=& 
\hat{p}_0\, {\mathcal H}
 + \hat{p}_0\hat{p}_{3}   \Big[ \frac{9}{2} - 4 \ln (\frac{4  \hat{p}_{3}^2}{\epsilon  y }) \Big]   
 + \frac{1}{4 {y }^2}{Y}_p( 2 - y^2-4y^{4}\nonumber \\
&&+3y^{6}-2\epsilon^{2}-2\epsilon^{4}+2\epsilon^{6}-4y^{2}\epsilon^{2}-5y^{2}\epsilon^{4} )   \, ,\nonumber \\
G_{-} &=&  {\mathcal H}+
 \hat{p}_{3} \Big[6 - 4 \ln (\frac{4 \hat{p}^2}{\epsilon  y }) \Big]
  +\frac{1}{y^{2}}{Y}_{p}( 1-y^{2}-2\epsilon^{2}+\epsilon^{4}-3y^{2}\epsilon^{2})\,, \nonumber \\
G_{0}&=&-6\,\hat{p}_{0}\, \hat{p}_{3} \ln \epsilon\,,
\end{Eqnarray}
where
\begin{eqnarray}
{\mathcal H}&=& 4 \hat{p}_0  \Big[\Li({\hat{p}_+})-\Li({\hat{p}_-}) 
- 2\,\Li(1 - \frac{{\hat{p}_-}}{{\hat{p}_+}})  
+ {{Y}_p} \ln (\frac{4\,y \,{{\hat{p}_3}}^2}{{{\hat{p}_+}}^2}) -{{Y}_w} \ln\epsilon\,\Big]\nonumber \\
&&
+2 Y_{w}(1-\epsilon^{2}) +\frac{2}{y^{2}}\hat{p}_{3}(1+y^{2}-\epsilon^{2})\ln\epsilon\,.
\end{eqnarray}
The polarized $\mathcal O(\alpha_{s})$ corrections are given by
\begin{eqnarray}
\label{totalpol}
\Gamma^{pol}_{NLO}&=&\Gamma_{loop}^{P}+\Gamma_{tree}^{P}\nonumber\\
&=&\frac{\alpha_{s} }{8 \pi^{2}}m_{t}C_{F} 
\bigg\{-3(a^{2}+b^{2})\hat{p}_{3}^{2}\ln\epsilon + a b\bigg[
  \frac{1}{4} \Big(-11+28y-16 y^{2} -8y^{3}+7y^{4}\nonumber\\
&& +\epsilon^{2}(4+8y -14y^{2})+7\epsilon^{4}\Big) 
+\Big(2-9y^{2}+y^{4}-\epsilon^{2}(4+3y^{2})+2\epsilon^{4}\Big)
\frac{\hat{p}_{3}}{y^{2}} Y_{p}
\nonumber \\
&&
+ 8\hat{p}_{3}^{2}\ln\Big(\frac{1-y}{(1-y)^{2}-\epsilon^{2}}\Big)
+\Big(3-3y^{2}+ 2\epsilon^{2}(4+y^{2})-2\epsilon^{4}\Big)\ln(\frac{1-y}{\epsilon})\nonumber \\
&&+\frac{8\hat{p}_{3}^{2}\, \hat{w}_{0}}{y^{2}} \ln\epsilon
+ 4\hat{p}_{0}\hat{p}_{3}\Big(
        2\Li (1-\frac{1-y}{\hat{p}_{-}})-2\Li (1-\frac{1-y}{\hat{p}_{+}})\nonumber\\
&&  -\Li (\hat{w}_{-})+\Li (\hat{w}_{+})+2 \ln (\frac{(1-y^{2})-\epsilon^{2}}{\epsilon^{2}})Y_{p} \Big)\nonumber \\
&&- \big(2+y^{4}-\epsilon^{2}(3+2y^{2})+\epsilon^{4}\big)\Big(2\Li(y)-\Li (\hat{w}_{-})-\Li (\hat{w}_{+})\Big)
\bigg]
\bigg\}\,.
\end{eqnarray}
 
\noindent As mentioned before the IR-- and mass singularities cancel in the sum of the
 one--loop and the tree--graph contributions.

 Next we discuss various limiting cases for the unpolarized and polarized
 rates, which, among others, serve to check on the correctness of our results.
 We have checked that our $ m_b \ne 0 $ results for the unpolarized rate
 agree with those given in \cite{Czar93(2)}.
 We do not, however, agree with the $ m_b \ne 0 $ results of \cite{Lial92}.

 The limit $ m_{H^+} \rightarrow 0 $ is of interest since, according to the
 Goldstone equivalence theorem, the unpolarized and polarized rates for
 $ t \rightarrow H^{+} + b $ become related to the unpolarized and polarized
 longitudinal rates of $ t \rightarrow W^{+} + b $ in the limit 
 $ m_{H^+}, m_{W^+} \rightarrow 0 $.
For model~1 and $ \tan \beta = 1 $ in the  $ m_{H^+} \rightarrow 0 $  limit one has

 \alpheqn
 \begin{Eqnarray} 
  \label{mHzerounpol} 
  \lim\limits_{m_{H^+} \rightarrow 0} \Gamma & = &
  \frac{m_t^3}{8 \pi} \frac{G_F}{\sqrt{2}}
  |V_{tb}|^2 (1 - \epsilon^2)^3 \bigg\{ 1 + \frac{\alpha_s}{\pi} C_F
  \frac{1 + \epsilon^2}{1 - \epsilon^2}\bigg[
  \frac{5 - 22 \epsilon^2 + 5 \epsilon^4}{4 (1 - \epsilon^4)} -
  2 \ln\epsilon \ln (1 - \epsilon^2)  \nonumber \\ 
& &-  2 \Li (1 - \epsilon^2)
- 2 \frac{1 - \epsilon^2}{1 + \epsilon^2} \ln \Big( \frac{1 - \epsilon^2}{\epsilon^2} \Big)
 -  \frac{4 - 5 \epsilon^2 + 7 \epsilon^4}{(1 - \epsilon^2) (1 - \epsilon^4)} \ln\epsilon  \bigg] \bigg\}\, ,\\[4mm]
  \label{mHzeropol} 
  \lim\limits_{m_{H^+} \rightarrow 0} \Gamma^{P} & = &
  \frac{m_t^3}{8 \pi} \frac{G_F}{\sqrt{2}}
  |V_{tb}|^2 (1 - \epsilon^2)^3 \bigg\{ 1 + \frac{\alpha_s}{\pi} C_F
  \frac{1}{1 - \epsilon^2}  \bigg[ - \frac{3}{4} (5 + \epsilon^2)  
- 2 (1 + \epsilon^2) \ln\epsilon   \ln (1 - \epsilon^2)\nonumber \\
&&  - 2 (1 - \epsilon^2) \ln \Big( \frac{1 - \epsilon^2}{\epsilon^2} \Big) 
  - \frac{4 + 5 \epsilon^2}{1 - \epsilon^2} \ln\epsilon 
+  (1 - 2 \epsilon^2) \Li (1 - \epsilon^2) \bigg] \bigg\}.
 \end{Eqnarray}
 \reseteqn

 \noindent We have checked that 
 expressions Eqs.~(\ref{mHzerounpol}) and (\ref{mHzeropol}) agree exactly with
 the $ m_{W^+} \rightarrow 0 $ limit of the corresponding longitudinal and
 polarized longitudinal rates in the process $ t \rightarrow W^+ + b $ listed
 in \cite{FGKM02}.
 This is nothing but the statement of the Goldstone equivalence theorem.
Our unpolarized result in 
 Eq.~(\ref{mHzerounpol}) agrees with the corresponding $ m_{H^+} \rightarrow 0 $
 result in \cite{Czar93(2)}.

 For the sake of comparison with results in the literature we take both 
 $ m_{H^+} \rightarrow 0 $ and 
 $ m_b \rightarrow 0 $ in Eqs.~(\ref{t2Hbtotalunpol}) and (\ref{totalpol}) without
 the cautionary proviso of keeping the term $m_b \tan \beta$ in model 2. In both
 models 1 and 2 one then obtains 

 \alpheqn
 \begin{Eqnarray} 
  \label{gbet}    
  \lim\limits_{m_{H^+} \rightarrow 0} \Gamma & \!=\! &
  \frac{m_t^3}{8 \pi} \frac{G_F}{\sqrt{2}} |V_{tb}|^2 \cot^2 \beta
  \left[ 1 \!+\! \frac{\alpha_s}{2\pi} C_F
  \left( \frac{5}{2} \!-\! \frac{2 \pi^2}{3} \right) \right], \nonumber\\
\\
  \lim\limits_{m_{H^+} \rightarrow 0} \Gamma^P & \!=\! &
  \frac{m_t^3}{8 \pi} \frac{G_F}{\sqrt{2}} |V_{tb}|^2 \cot^2 \beta
  \left[ 1 \!-\! \frac{\alpha_s}{2\pi} C_F
  \left( \frac{15}{2} \!-\! \frac{\pi^2}{3} \right) \right]\nonumber\\ 
 \end{Eqnarray}
 \reseteqn

 \noindent which, when setting $ \cot \beta = 1 $, agree
 exactly with Eqs.~(48) and (49) of \cite{FGKM02} and the $\epsilon \to 0$ limits
 of Eqs.~(\ref{mHzerounpol}) and (\ref{mHzeropol}).
 The unpolarized rate in this limit agrees with the
 corresponding results in \cite{Czar93(1),Czar93(2),LiYa92}.

 When $ m_{H^+} $ approaches $ m_t $ for $ m_b \rightarrow 0 $ one has

 \alpheqn
 \begin{Eqnarray} 
  \label{limit}   
  \lim\limits_{m_{H^+} \rightarrow m_t}
  \frac{\Gamma}{\Gamma_{Born}} & = &
  1 + \frac{\alpha_s}{2 \pi} C_F \left[\frac{13}{2}  - \frac{4 \pi^2}{3} -
  3 \ln (1 - y^2) \right], \nonumber\\ \\
  \lim\limits_{m_{H^+} \rightarrow m_t}
  \frac{\Gamma^P}{\Gamma_{Born}^P} & = &
  1 + \frac{\alpha_s}{2 \pi} C_F \left[ 1 - \pi^2 -
  3 \ln (1 - y^2) \right]\,\nonumber \\
 \end{Eqnarray}
 \reseteqn

 \noindent where the $\tan \beta$ dependence has disappeared by dividing out the 
 relevant Born terms.
 For the unpolarized rate the limiting expression
 agrees with the corresponding limit given in \cite{Czar93(2)}.

 Finally, we consider the limit $ m_b \rightarrow 0 $
 keeping the charged Higgs mass finite.
 This results in very compact expressions for the
 unpolarized and polarized rates.
 Due to the smallness of the bottom quark mass and
 the fact that the bottom quark mass corrections are of
 $ O(m_b^2 / m_t^2) $ the $ m_b \rightarrow 0 $ formulae
 give quite good approximations to the exact formulae
 for Higgs masses as long as the Higgs mass is not close to the top quark mass.

 One obtains

 \alpheqn
 \begin{eqnarray} 
 \label{mbzerounpol} 
   \lim\limits_{m_{b}\to 0}\Gamma & = &
   \frac{m_t}{16 \pi}(1 - y^2)^2 (a^2 + b^2)\bigg\{
   1 + \frac{a^2 - b^2}{a^2 + b^2}\frac{2 \epsilon}{1 - y^2}  
+ \frac{\alpha_s}{2\pi} C_F
   \bigg[ \frac{9}{2} - \frac{2 \pi^2}{3}
-  \frac{4 y^2}{1 - y^2} \ln y \nonumber\\
& & 
  +
   \Big( \frac{2 - 5 y^2}{y^2} -
   4 \ln y \Big) \ln (1 - y^2)  - 4 \Li (y^2) 
+ \frac{(a - b)^2}{a^2 + b^2} 3 \ln \epsilon \bigg] \bigg\} \,,\\[2mm]
%
   \label{mbzeropol}
   \lim\limits_{m_{b}\to 0}\Gamma^P & = &
   \frac{m_t}{16 \pi}(1 - y^2)^2 2 a b \bigg\{
   1 + \frac{\alpha_s}{2\pi} C_F  \bigg[ -
   \frac{11 - 6 y - 7 y^2}{2 (1 + y)^2} +
   \frac{1 + 2 y^2}{(1 - y^2)^2} \frac{\pi^2}{3}
   \nonumber \\ 
& &+\,  \frac{2 - 9 y^2 + y^4}{(1 - y^2) y^2} \ln (1 + y)  
+ \frac{2 - 5 y^2}{y^2} \ln (1 - y)-   4 \Li (y) 
   \nonumber \\
& & +\,  \frac{8 + 4 y^4}{(1 - y^2)^2} \Li (-y)
 - \frac{(a - b)^2}{2 a b} 3 \ln \epsilon \bigg] \bigg\}.
 \end{eqnarray}
 \reseteqn
 Again we have checked that the $m_b \to 0 $ limit of the unpolarized rate agrees 
 with the corresponding result of \cite{Czar93(2)}. Note that the seemingly mass 
 singular terms proportional to $ \ln \epsilon $ in Eq.~(\ref{mbzerounpol}) and 
 (\ref{mbzeropol}) are not in fact mass singular, since they are multiplied by the 
 factor $ ( a - b )^2 $, which is proportional to $ m_b^2 $ in both models~1 and 2.
 Although the contributions proportional to $(\epsilon^2 \ln \epsilon)$ formally
 vanish for $ m_b \rightarrow 0 $, they can become numerically quite large for 
 $ m_b = 4.8 $ GeV in model~2, depending, of course, on the value of $ \tan \beta $.
 This can be seen by calculating the ratios of the coupling factor expressions that 
 multiply the $ \ln \epsilon $ term in Eqs.~(\ref{mbzerounpol}) and (\ref{mbzeropol}).
 In model 1 one has  $ (a-b)^2 / (a^2+b^2) = 2 \epsilon^2 /(1 + \epsilon^2) $ and
 $ (a-b)^2 / (2ab) = 2 \epsilon^2 /(1 - \epsilon^2) $, whereas in model 2 one has
 $ (a-b)^2 / (a^2+b^2) = 2 \epsilon^2 \tan ^4 \beta /(1 + \epsilon^2 \tan ^4 \beta) $ 
 and $ (a-b)^2 / (2ab) = 2 \epsilon^2 \tan ^4 \beta /(1 - \epsilon^2 \tan ^4 \beta) $.
 In model 1 the $ \ln \epsilon $ contribution is negligible but not in model 2. 
 For example, in model 2 one finds $ (a - b)^2 / (a^2 + b^2) = 1.77 $ and
 $ (a - b)^2 / (2ab)= - 2.31 $ for $ \tan \beta = 10 $. With a little bit of algebra 
 one finds that the NLO 
 corrections in model~2 are in fact dominated by the $(\epsilon^2 \ln \epsilon)$ 
 contributions for larger values of $ \tan \beta $ as also noted in \cite{Czar93(2)}.
 This is evident in Figs.~4a and 4b where the radiative corrections to the
 Born term rate can be seen to be as large as $ - 50 \% $ compared
 to the $ \approx - 10 \% $ expected from the corresponding corrections
 in the decay $ t \rightarrow W^{+} + b $ \cite{FGKM02}. 
Note, though, that the radiative corrections become much smaller if one uses the running 
bottom quark mass $\bar{m}_{b}(m_{t})=1.79$~GeV, i.e. $\epsilon^{2}\ln\epsilon=-0.00048$
instead of the fixed pole mass $m_{b}=4.8$~GeV, i.e. $\epsilon^{2}\ln\epsilon=-0.00271$. 

 We now turn to a more detailed discussion of our numerical results on the 
 radiative corrections to the unpolarized rates and the polarization asymmetry in model 1 and 2.

 We start our discussion with model 1. As input values for our numerical evaluation 
 we use $ m_b = 4.8 \,{\rm GeV} $ and $ m_t = 175 \,{\rm GeV} $ 
as in the case of Born term.
 The strong coupling constant is evolved from $ \alpha_s(M_Z) = 0.1175 $ to
 $ \alpha_s(m_t) = 0.1070 $ using two-loop running. Fig.~2a shows that the radiative 
 corrections lower the model 1 rate by approximately $10\%$ over the whole range of
 $ y = m_{H^+} / m_t $. The radiative corrections in Fig.~2b are of similar
 size but cannot be discerned at the scale of the figure. The radiative corrections 
 to the polarized rate are of similar size and go in the same direction, which means 
 that the radiative corrections to the polarization asymmetry are quite small.
 This can be seen in Fig.~3, which shows that the radiative corrections to the 
 polarization asymmetry are indeed quite small and lower $ \alpha_H $ only by 
 $\approx 2 \permille$. No value of $ \tan \beta $ is given in Fig.~3, since 
 in model 1 the polarization asymmetry does not depend on $ \tan \beta $ at LO 
 and NLO, and, in fact, at any order in $\alpha_{s}$. 

 Figs.~4a and 4b show that in model~2
 the radiative corrections are substantial, which is mostly due to the
 $ \epsilon^2 \ln \epsilon $ contribution discussed above. They reduce the LO
 rate by $\approx 50\%$ over much of the shown $y$ and $\tan \beta$ ranges.
 The barely visible dotted curves in Figs.~4a and 4b are drawn using the 
 ``kinematical''
 $ m_b \rightarrow 0 $ approximations Eqs.~(\ref{bornzerobmass1}) and
 (\ref{bornzerobmass2}) (LO) and
 Eqs.~(\ref{mbzerounpol}) and (\ref{mbzeropol}) (NLO).
 In these equations the bottom
 quark mass has been set to zero whenever the scale of $ m_b $ is set by
 $ m_t $ as in the kinematical factors and not by $ m_t \cot^{2} \beta $
 as in the coupling factors.
 As Fig.~4 shows, the ``kinematical'' $ m_b \rightarrow 0 $
 approximation is an excellent approximation for both the unpolarized and
 polarized rate.

 In Fig.~5 we show that in model~2
 the radiative corrections to the LO Born term result are
 substantial and reduce the size of the polarization asymmetry
 by $ \approx 25 \% $ over much of the range of the Higgs mass.
 The $ m_b \rightarrow 0 $ approximation is quite good except in the endpoint
 region. At the scale of
 Fig.~5 the $ m_b \ne 0 $ corrections are barely visible.
 In Fig.~5b we fix the mass of the charged Higgs boson at
 $ m_{H^+} = 120 \,{\rm GeV}$ and vary $ \tan \beta $
 between $ 0 $ and $30$.
 The LO zero position $ \tan \beta = \sqrt{m_t/m_b}=6.04 $ is shifted
 upward by approximately one unit by the radiative corrections. The radiative 
 corrections can be seen to become quite small for the larger $\tan \beta$ values.
 The radiative corrections are
 largest around the zero position of $ \tan \beta $ at
 $ \tan \beta  \approx 7 $.


\section{Concluding remarks}

 We have calculated the $ O(\alpha_s) $ radiative corrections to polarized
 top quark decay into a charged Higgs and a bottom quark in two variants
 of the two-Higgs-doublet model.
 We have checked our unpolarized results against
 known results and found agreement.
 Using the same techniques we have calculated the polarized rate.
 Further, we have compared our polarized results with
 the corresponding polarized results in the decay $ t \rightarrow W^{+} + b $
 appealing to the Goldstone equivalence theorem.
 Because of our numerous cross-checks we are quite confident that our
 new results on the polarized rates are correct.
 We have found very compact $ O(\alpha_s) $ expressions for
 the unpolarized and polarized rates in $m_{b}=0$ limit, 
 which can be usefully employed to scan
 the predictions of the 2HDM $ (m_{H^+},\tan \beta) $ parameter space.
 
 We have found that a measurement of the polarization asymmetry in
 the decay of a polarized top quark into a charged Higgs and a bottom quark
 $t(\uparrow) \to b + H^+$ can discriminate between the two variants of the 
 two-Higgs-doublet model investigated in this paper. In model 1 the polarization 
 asymmetry does not depend on $\tan \beta$ and stays very close to $\alpha_H=1$ 
 for a large range of charged Higgs mass values. 
This is different in model 2, in which the polarization asymmetry strongly depends
 on the values of $\tan \beta$ and the charged Higgs mass and can vary between
 $\alpha_H=1$ and $\alpha_H=-1$ depending on the parameter values.
 The radiative corrections to the polarization asymmetry are quite small 
($\approx 2 \permille$) in model 1. Again this is different in model~2 where the 
 radiative corrections to the polarization asymmetry are important 
 and can become as large as $25 \%$.


\vspace{1truecm} \noindent {\bf Acknowledgements:}
 A. Kadeer acknowledges the support of the DFG (Germany) through the
 Graduiertenkolleg ``Eichtheorien'' at the University of Mainz.
 M.~C. Mauser was partly supported by the DFG (Germany) through the
 Graduiertenkolleg ``Eichtheorien'' at the University of Mainz and
 by the BMBF (Germany) under contract 05HT9UMB/4.
 M.~C. Mauser would also like to thank K. Schilcher for his support.

\begin{appendix}
\section{\label{loopInt-t2Hb} Loop integrals}

A detailed discussion of the one--loop integrals can be found 
in e.g. Sec.~4 of \cite{Denn93}.

The two--point one--loop scalar 
integral\footnote{Note that $q=p_{t}-p_{b}$, $q^{2}=m_{H^{+}}^{2}$ and $m_{t}>m_{b}+m_{H^{+}}$\,.}:
\begin{Eqnarray}
B_{0}(q,m_{t},m_{b} )&:&=\frac{\mu^{4-D}}{i\pi^{2}}\int\!\!\frac{d^Dk}{(2\pi)^{D-4}}
\frac{1}{\left[\left(p_{t}-k\right)^2-m_t^2\right]
\left[\left(p_b-k\right)^2-m_b^2\right]}\nonumber  \\
&=& \frac{1}{\omega} -\gamma_{E}+2+\ln\frac{4\pi\mu^{2}}{m_{t}^{2}}   + \frac{1-y^{2}-\epsilon^{2}}{y^{2}}\ln\epsilon
    + \frac{2\hat{p}_{3}}{y^{2}}Y_{p}\,.
\end{Eqnarray}

The three--point one--loop scalar integral:
\begin{Eqnarray}
&& C_0(q,m_{t},m_{t}, m_{b}, m_{g})
:=\frac{\mu^{4-D}}{i\pi^{2}}\int\frac{d^D k}{(2\pi)^{D-4}}
\frac{1}{\left[\left(p_b-k\right)^2-m_b^2\right]
\left[\left(p_t-k\right)^2-m_t^2\right]\left(k^2-m_{g}^{2}\right)}\nonumber\\
&& =  -\frac{1}{m_{t}^{2}}\frac{1}{2\,{\hat{p}_3}}
\bigg\{
      \left( \ln\epsilon +{Y}_p  \right)  \left(  {Y}_p  + 2 {Y}_w  \right)+ 
      \left( \ln\epsilon - 2 \ln\Lambda \right)  {Y}_{p}  -\Li(1 - \frac{{\hat{w}_-}}{{\hat{w}_+}}) + 
      \Li(1 - \frac{{\hat{p}_-} {\hat{w}_-}}{{\hat{p}_+} {\hat{w}_+}})   \bigg\}\,.\nonumber\\ 
\end{Eqnarray}

The three--point one--loop vector integral:
\begin{Eqnarray}
C^\mu(q,m_{t},m_{t}, m_{b}, m_{g})&:=&\frac{\mu^{4-D}}{i\pi^{2}}
\int\frac{d^D k}{(2\pi)^{D-4}}
\frac{k^{\mu}}{\left[\left(p_b-k\right)^2-m_b^2\right]
\left[\left(p_t-k\right)^2-m_t^2\right]\left(k^2-m_{g}^{2}\right)}\nonumber\\
&=&C_{1}\, p_{t}^{\mu}+ C_{2}\, p_{b}^{\mu} \,.
\end{Eqnarray}
where 
\begin{Eqnarray}
\label{C1C2loop}
C_{1} &=&  \frac{1}{m_{t}^2} \frac{1}{y^2}
       \left[ \ln\epsilon + \left(1-y^{2}-\epsilon^{2} \right)\frac{{Y}_{p} }{2\hat{p}_{3}} \right]\,, \\[4mm]
C_{2}&=&                                                                                                                                 -\frac{1}{m_{t}^2} \frac{1}{y^2}
      \left[ \ln\epsilon + \left(1+y^{2}-\epsilon^{2} \right)\frac{{Y}_{p} }{2\hat{p}_{3}} \right]\,.
\end{Eqnarray} 

\section{Integration of the soft gluon factor}
\label{SGF-int}
\begin{Eqnarray}
S(\Lambda) &:=& (-4\pi
m_{t})\frac{1}{2m_{t}}\frac{1}{(2\pi)^{5}}\int \frac{d^{3}\vec{p}_{b}}{2E_{b}}
     \frac{d^{3}\vec{q}}{2E_{H}}
     \frac{d^{3}\vec{k}}{2E_{g}} \,\delta^{(4)}(p_{t}-q-p_{b}-k) |{\cal
M}|^{2}_{SGF} \nonumber \\
&=& \frac{\alpha_{s}}{4\pi}C_{F}\Bigg(4 {\hat{p}_3} \Big[ \ln (\frac{4
{{{\hat{p}}_3}}^2}{\epsilon  y \Lambda }) -2\Big]
    - 2 (1-\epsilon^{2})Y_w + 2 {\epsilon }^2\,Y_p + 2 {{\hat{p}}_0}
  \Big\{
   2 \Li(1 - \frac{{{\hat{p}}_-}}{{{\hat{p}}_+}})  \nonumber\\
&&  -
     \Li(1 - \frac{{{\hat{w}}_-}}{{{\hat{w}}_+}})
     + \Li(\frac{2 {{\hat{p}}_3}}{{{\hat{p}}_+} {{\hat{w}}_+}}) 
- {Y}_p \Big[ 2\ln (\frac{4 {{{\hat{p}}_3}}^2}{{{\hat{p}}_+}
{{\hat{w}}_+}\Lambda}) -Y_{p}+1\Big]
     \Big\}\Bigg)\,,
\end{Eqnarray}

\noindent where the soft gluon  (or eikonal) factor 
$|{\cal M}|^{2}_{SGF}$ is defined in Eq.~(\ref{MsgfDef}).
\section{\label{basicInt}Basic integrals for the tree--graph phase space integrations}
Details of the calculation of basic integrals can be found in \cite{FGKM02}.
Note that our notation differs from the notation in \cite{FGKM02}.
{\allowdisplaybreaks
\begin{Eqnarray}
\label{R4t2Hb}
&& R(-1)= -2 {{ \hat{p}}_3} + 2 {{ \hat{p}}_3}\ln (\frac{2 {{ \hat{w}}_+} {{ \hat{p}}_3}}{{y }^2\epsilon_{2}} )
           - \left( 1 - {\epsilon }^2 + {y }^2 \right)  { {Y}_w}   \,,\\[2mm]
&& R(0)=  \frac{1}{2} \ln \Big[\frac{( 1 - y )^2-\epsilon^2 }{( 1 + y )^2  
         -{\epsilon }^2  }\Big] + \ln (\frac{{{ \hat{w}}_+}}{y\epsilon_{2} })   \,,\\[2mm]
&& R(0,0)= {\left( 1 - y  \right) }^2-{\epsilon }^2 \,,\\[2mm]
&& R(0,-1)={{ \hat{p}}_3} \left( 1 - {\epsilon }^2 + {y }^2 \right)   - 2 {y }^2 { {Y}_w}   \,,\\[2mm]
&& R(-1,0)= 2 \ln (\frac{1 - y }{\epsilon }) \,,\\[2mm]
&& R(-1,-1)= -2 {{ \hat{p}}_3} + 2\left( 1 - {y }^2 \right)  { {Y}_p}-2{y }^2 { {Y}_w}   \,,\\[2mm]
&& R(-2,0)= {\epsilon }^{-2} - {\left( 1 - y  \right) }^{-2}   \,,\\[2mm]
&& R(-2,-1)= \frac{2 {{ \hat{p}}_3}}{{\epsilon }^2} - 
             \frac{2}{1 \!-\! {y }^2}\Big[ \left( 1 \!+\! {y }^2 \right)  { {Y}_p} 
             + {y }^2 { {Y}_w} \Big]\,,\\[2mm]
&& R(1,0)= \frac{{\left( 1 - y  \right) }^4}{2}-\frac{{\epsilon }^4}{2}   \,,\\[2mm]
%
\label{S4t2Hb}
&& S(0)=\Li({{ \hat{w}}_+}) -\Li({{ \hat{w}}_-}) - 2 \Li(1 - \frac{{{ \hat{p}}_-}}{{{ \hat{p}}_+}})    + 2\ln (\frac{2 {{ \hat{w}}_+} {{ \hat{p}}_3}}{{y }^2 \epsilon_{2}})
            { {Y}_p} - 2 {{ {Y}_p}}^2 + 2{Y}_{w} \ln\epsilon   \,,\\[2mm]
&& S(1)= \frac{1}{ {{ \hat{p}}_3}}\Big\{ \Li(- \frac{y }{{{ \hat{w}}_+}} ) - 
         \Li(- \frac{y  {{ \hat{p}}_-}}{{{ \hat{p}}_+} {{ \hat{w}}_+}} )  + 
          \ln (\frac{2 {{ \hat{p}}_3}}{y \epsilon_{2}})  { {Y}_p} - {{ {Y}_p}}^2 \Big\} \,,\\[2mm]
&& S(0,-1)=-\frac{1}{4}\big[ (1 -y)^{2}- \epsilon^{2} \big]  
         \big[ {\epsilon }^2 -\left(3 - y  \right)  \left( 1 + y  \right)  \big] 
- \left(1- {y }^4 \right)  \ln (\frac{1 - y }{\epsilon }) \nonumber\\ 
&&\qquad\qquad\quad  
           + 2 {y }^2 \Big[2 \Li(y ) - \Li({ \hat{w}_-}) - \Li({ \hat{w}_+}) \Big]
+ 2 \left( 1 - {\epsilon }^2 + {y }^2 \right)  { \hat{p}_3} { {Y}_p}\,,\\[2mm] 
&& S(0,0)= 2 \left( {{ \hat{p}}_3} - {\epsilon }^2 { {Y}_p} - {y }^2 { {Y}_w} \right) \,,\\[2mm]
&& S(0,1)= \Li({{ \hat{w}}_-}) + \Li({{ \hat{w}}_+})-2 \Li(y ) \,,\\[2mm]
&& S(1,0)= \frac{{{ \hat{p}}_3}}{2}\left( 1 + {\epsilon }^2 + 5 {y }^2 \right) 
           - {\epsilon }^4 { {Y}_p} 
           - {y }^2 \left( 2 + {y }^2 \right)  { {Y}_w} \,,\\[2mm]
&& S(1,1)= {\epsilon }^2 - {\left( 1 - y  \right) }^2 
        + 2 \left( 1 - {y }^2 \right)  \ln (\frac{1 - y }{\epsilon })- 4 {{ \hat{p}}_3} { {Y}_p}\nonumber \\ 
      &&\qquad\qquad + 
      \left( 1 + {y }^2 \right)  \Big[  \Li({{ \hat{w}}_-}) + \Li({{ \hat{w}}_+})-2 \Li(y )
         \Big]  \,,\\[2mm]
&& S(2,1)= \frac{1}{4} \big[{\epsilon }^2 - {\left( 1 - y  \right) }^2 \big]  
           \big[ 4 + {\epsilon }^2 + {\left( 1 - y  \right) }^2 + 8 {y }^2 \big]  
+ 3 \left( 1 - {y }^4 \right)  \ln (\frac{1 - y }{\epsilon })\nonumber   \\
      && \qquad\qquad -  \left( 1 + 4 {y }^2 + {y }^4 \right)  \Big[ 2 \Li(y ) - \Li({ \hat{w}_-}) -
         \Li({ \hat{w}_+}) \Big]\nonumber   \\
      && \qquad\qquad  - 2{ \hat{p}_3} \left( 3 + {\epsilon }^2 
            + 3 {y }^2 \right)   { {Y}_p} \,.
\end{Eqnarray}
}

\end{appendix}


\begin{thebibliography}{99}
  \bibitem{MaPa97} G.~Mahlon and S.~Parke, Phys.~Rev.~{\bf D55} (1997) 7249.
  \bibitem{EsMa02} D.~Espriu and J.~Manzano, {\bf hep-ph}/0209030.
  \bibitem{PaSh96} S.~Parke and Y.~Shadmi, Phys.~Lett.~{\bf B387} (1996) 199.
  \bibitem{KoMePr02} M.~Fischer, S. Groote, J.G.~K\"orner, M.C.~Mauser and B. Lampe,
          Phys.~Lett. {\bf B451} (1999) 406.
  \bibitem{Czar93(1)} A.~Czarnecki and S.~Davidson,
  Phys.~Rev.~{\bf D47} (1993) 3063.
  \bibitem{Czar93(2)} A.~Czarnecki and S.~Davidson,
  Phys.~Rev.~{\bf D48} (1993) 4183.
  \bibitem{Lial92} C.S.~Li et al.,
  Phys.~Lett.~{\bf B285} (1992) 137.
  \bibitem{LiYa92} J.~Liu and Y.P.~Yao,
  Phys.~Rev.~{\bf D46} (1992) 5196.
  \bibitem{RTLS91} J.~Reid, G.~Tupper, G.~Li, and M.S.~Samuel,
  Z.~Phys.~{\bf C51} (1991) 395.
  \bibitem{LiYu90} C.S.~Li and T.C.~Yuan,
  Phys.~Rev.~{\bf D42} (1990) 3088.
%
  \bibitem{LiYu90-2} C.S.~Li and T.C.~Yuan,
  Phys.~Rev.~{\bf D47} (1993) 2156(E).
  \bibitem{LiYa90} J.~Liu and Y.P.~Yao,
  Report~No.~{\bf UM-TH-90-09} (1990) (unpublished).
%
  \bibitem{LiYa90-2} J.~Liu and Y.P.~Yao,
  Int.~J.~Mod.~Phys.~{\bf A6} (1991) 4925.
  \bibitem{FGKM02} M.~Fischer, S.~Groote, J.G.~K\"orner,
  and M.C.~Mauser, \\[2mm]
  Phys.~Rev.~{\bf D65} (2002) 054036,
  {\bf hep-ph}/0101322.
  \bibitem{GuHaKaDa90}
  J.F.~Gunion, H.E.~Haber, G.L.~Kane and S.~Dawson, \\[2mm]
  {\it The Higgs Hunter's Guide'}, (Addison-Wesley, Reading, MAA, 1990).
\bibitem{pdg06} Particle Data Group, W.~M.~Yao {\it et al}, Journal of Physics 
{\bf G33}~(2006)~1.
  \bibitem{BrLe80} E.~Braaten and J.P.~Leveille,
  Phys.~Rev.~{\bf D22} (1980) 715.
  \bibitem{Denn93} A.~Denner,
  Fortschr.~Phys.~{\bf 41} (1993) 307.

 \end{thebibliography}
\end{document}